\documentclass[prl,superscriptaddress,preprint,endfloats]{revtex4}
\newcommand {\FS}{Fe$_3$Sn$_2${}}
\newcommand {\ECS}{EuCd$_2$Sb$_2${}}

\newcommand {\rhoyx}{$\rho _{yx}${}}

\newcommand {\rhoyxAHE}{$\rho _{yx,{\mathrm{AHE}}}$}

\newcommand {\Bin}{$B_{\mathrm{in}}${}}

\newcommand {\Bsatin}{$B_{\mathrm{sat,in}}$}

\newcommand {\TN}{$T_{\mathrm{N}}$}

\usepackage[dvips]{graphicx}
\usepackage{mediabb}
\usepackage{bm} 
\usepackage{pifont}
\usepackage{amsmath, amsthm, amssymb}
\usepackage{color}
\usepackage{dcolumn} 
\usepackage{amsmath, amsthm, amssymb, pifont, wasysym}
\usepackage[utf8]{inputenc}

\definecolor{mygray}{gray}{0.5}

\begin{document}
\title{Magneto-cubic and magneto-linear dependence \\
	observed in an in-plane anomalous Hall magnet}
\author{Ayano Nakamura}
\affiliation{Department of Physics, Institute of Science Tokyo, Tokyo 152-8551, Japan}
\author{Shinichi Nishihaya}
\affiliation{Department of Physics, Institute of Science Tokyo, Tokyo 152-8551, Japan}
\author{Mitsuru Akaki}
\affiliation{Institute for Materials Research (IMR), Tohoku University, Sendai 980-8577, Japan}
\author{Motoi Kimata}
\affiliation{Advanced Science Research Center, Japan Atomic Energy Agency (JAEA), Ibaraki, 319-1195, Japan}
\author{Kenta Sudo}
\affiliation{Institute for Solid State Physics (ISSP), The University of Tokyo, Kashiwa 277-8581, Japan}
\author{Yuki Deguchi}
\affiliation{Department of Physics, Institute of Science Tokyo, Tokyo 152-8551, Japan}
\author{Hsiang Lee}
\affiliation{Department of Physics, Institute of Science Tokyo, Tokyo 152-8551, Japan}
\author{Tadashi Yoneda}
\affiliation{Department of Physics, Institute of Science Tokyo, Tokyo 152-8551, Japan}
\author{Masaki Kondo}
\affiliation{Institute for Solid State Physics (ISSP), The University of Tokyo, Kashiwa 277-8581, Japan}
\author{Hiroaki Ishizuka}
\affiliation{Department of Physics, Institute of Science Tokyo, Tokyo 152-8551, Japan}
\author{Masashi Tokunaga}
\affiliation{Institute for Solid State Physics (ISSP), The University of Tokyo, Kashiwa 277-8581, Japan}
\author{Masaki Uchida}
\email[Author to whom correspondence should be addressed: ]{m.uchida@phys.sci.isct.ac.jp}
\affiliation{Department of Physics, Institute of Science Tokyo, Tokyo 152-8551, Japan}
\affiliation{Toyota Physical and Chemical Research Institute (TPCRI), Nagakute 480-1192, Japan}

\begin{abstract}
\textbf{The Hall effect, particularly that arising from in-plane magnetic field, has recently emerged as a sensitive probe of quantum geometric properties in solids. Especially in trigonal systems, in-plane anomalous Hall effect (AHE) can be explicitly induced by nontrivial off-diagonal coupling between the magnetic field and the Hall vector on the principal plane. Here we elucidate multipolar dependence of the off-diagonal coupling in the in-plane AHE, by systematically measuring on the (001) principal plane of trigonal antiferromagnet {\ECS} thin films for each magnetic phase. Around zero field, magneto-cubic dependence of anomalous Hall resistivity is clearly observed not only in the paramagnetic phase but also even in the antiferromagnetic phase. An off-diagonal component of the octupolar tensor also exhibits unconventional decay above the magnetic ordering temperature, roughly depending on the inverse temperature to the third power. In the forced ferromagnetic phase, on the other hand, magneto-linear dependence dominantly appears and notably persists up to very high fields. Our findings clarify key aspects of the off-diagonal coupling in the in-plane AHE, paving the way for its future investigations and potential applications beyond conventional expectations about the Hall effect.}
\end{abstract}
\maketitle

The Hall effect has long served as a fundamental tool for probing the interplay between  conduction and magnetism in solids \cite{Hall1, Hall2, AHEreview}. Beyond its classical manifestation due to the Lorentz force, the anomalous Hall effect (AHE) has emerged as a key transport phenomenon in magnetic materials, arising from mechanisms such as spin-orbit coupling and asymmetric scattering. In recent years, significant progress has been made in understanding the intrinsic contribution to AHE, which are now firmly established as manifestation of the Berry curvature of Bloch electrons \cite{AHEreview, AHEexp1, AHEexp2}. On the other hand, in-plane AHE - where the Hall voltage emerges within the plane defined by the magnetization and current as illustrated in Fig. 1\textit{A} - has attracted growing interest due to its unconventional origins \cite{iAHEthe1_Q, iAHEthe2_Q, iAHEthe3_Q, iAHEthe4_Q, iAHEthe7_Q, iAHEthe8, iAHEthe5, iAHEthe6, iAHEthe7}. Unlike the conventional AHE observed in out-of-plane magnetized systems, the in-plane AHE may not require net spin magnetization perpendicular to the Hall deflection plane. Instead, it is associated with pure orbital magnetization induced under appropriate symmetry conditions. 
Nowadays, a series of experimental observation of in-plane AHE, where three-fold symmetric signal for the in-plane field rotation is confirmed isolated from the out-of-plane AHE, has emerged, fueled by a growing number of reports in various materials. Examples include magnetic Weyl semimetals \cite{ECS_iAHE, Fe3Sn2_iAHE, EZS_iAHE, Co3Sn2S2_iAHE}, a nonmagnetic Dirac semimetal  \cite{CA_iAHE}, and a ferromagnetic oxide  \cite{SRO_iAHE}. These experimental studies have also proposed a variety of origins for this phenomenon, such as an orbital magnetization model \cite{ECS_iAHE} and an anomalous orbital polarizability model \cite{Fe3Sn2_iAHE}. 

In contrast to the conventional out-of-plane AHE, in-plane AHE is constrained not only by time-reversal symmetry breaking but also by various crystalline symmetries, more precisely those of the magnetic point group.
For example, for emergence of the in-plane AHE, the system must lack both out-of-plane $C_{2n}$ rotational axes and an in-plane mirror plane \cite{iAHE_symmetry}. Low-symmetry systems without any rotational axes perpendicular to the Hall deflection plane are a reasonable candidate, where the in-plane Hall response with one-fold symmetry with respect to the in-plane field rotation can be expected. In this case, however, it is practically difficult to distinguish it from other extrinsic effects such as anisotropy in the ordinary Hall effect \cite{IrO2_Hall, RuO2_Hall1, RuO2_Hall2}.
In contrast, rather high-symmetry systems with $C_3$ rotational axis perpendicular to the plane, such as  trigonal (001) and cubic (111) planes, are a more compelling candidate, displaying distinct  three-fold symmetry for the in-plane field rotation. Notably, in trigonal systems, a finite Hall response occurs on the principal plane while it is not the case in isotropic cubic systems. This indicates that intrinsic off-diagonal coupling between the applied magnetic field and the Hall vector can be studied through the in-plane AHE in trigonal systems.

In previous experimental observations of in-plane AHE on the trigonal (001) principal plane, a clear three-fold symmetric (3$\varphi$) signal with respect to the in-plane field angle $\varphi$ has been demonstrated for {\ECS} and {\FS} \cite{ECS_iAHE, Fe3Sn2_iAHE}. However, its detailed in-plane field dependences involving the intrinsic off-diagonal coupling are still elusive. In trigonal systems that belong, for example, to the point group $\bar{3}m$ as in {\ECS} and {\FS}, the magneto-linear coupling ($\propto B$) to {\rhoyx} is prohibited due to their symmetry and the magneto-cubic coupling ($\propto B^3$) serves as the leading term \cite{iAHE_symmetry}. Considering magnetic ordering at the ground state, however, the magneto-linear coupling can be a leading term, for example, in the in-plane antifferomagnetic (AFM) state belonging to the magnetic point group 2/$m$1', as compared in Fig. 1\textit{B}. In this case, the 3$\varphi$ dependence can still appear, if the ordered spins also rotate for the in-plane field rotation. In the orbital magnetization model, in-plane AHE observed in {\ECS} has be interpreted as manifestation of out-of-plane Weyl points splitting and orbital magnetization induced by the in-plane spin magnetic field, where the order of off-diagonal coupling is affected by the order $n$ in a relativistic cross term such as $k_y^{n}\sigma_z$ between the in-plane momentum and the out-of-plane spin operator \cite{ECS_iAHE}. 
In the anomalous orbital polarization model, on the other hand, in-plane AHE observed in {\FS} have been explained by the in-plane Berry connection change by the in-plane orbital magnetic field, where the magneto-linear coupling can be also expected \cite{Fe3Sn2_iAHE}.

In this study, we have clarified the magneto-cubic and magneto-linear behavior of in-plane AHE by measuring its detailed field and temperature dependences  in {\ECS} thin films.
{\ECS} crystallizes in a trigonal crystal structure with space group $P\bar{3}m1$, where Eu triangular lattices and Cd$_2$Sb$_2$ blocks are alternately stacked along the out-of-plane $c$-axis. 
{\ECS} is a suitable  system for studying in-plane AHE in terms of crystal symmetry and electronic structure.
The {\ECS} (001) principal plane has an out-of-plane $C_3$ rotational axis but not an in-plane mirror plane, satisfying the symmetry conditions required for in-plane AHE.
Also, Cd $5s$ and Sb $5p$ bands are thought to be inverted in {\ECS}, leading to a simple yet important band structures, which hosts a few pairs of Weyl points near the Fermi energy under the field \cite{ECS_Weyl}. This generates substantial Berry curvatures within the relatively small Fermi surface \cite{ECS_film, ECS_Weyl, ECS_nMC}, resulting in very large AHE under the out-of-plane magnetic field \cite{ECS_film, ECS_nMC}.

\begin{figure*}
	\begin{center}
		\includegraphics[width=14cm]{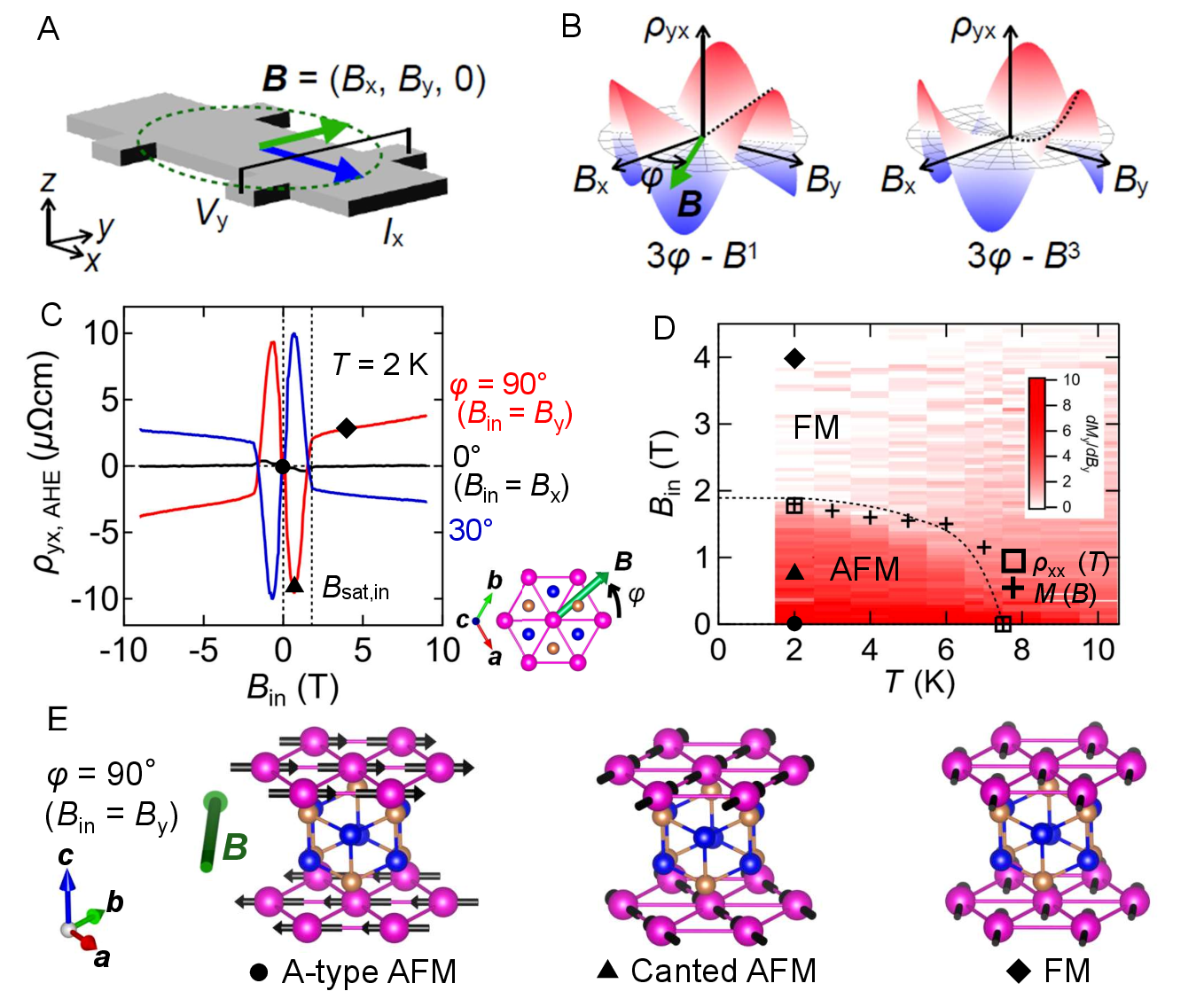}
		\caption{In-plane AHE and its emergence in {\ECS}. (\textit{A}) Schematic configuration of in-plane AHE, where transverse voltage $V_y$ is induced in response to electric current $I_x$ under in-plane magnetic field $\boldsymbol{\mathit{B}}$. (\textit{B}) Two possible types of in-plane field dependence ($B^1$ or $B^3$) in the Hall resistivity {\rhoyx}, both showing three-fold symmetry (3$\varphi$) for the in-plane field rotation with azimuthal angle $\varphi$. (\textit{C}) Anomalous Hall resistivity {\rhoyxAHE} of {\ECS}, appearing under the in-plane magnetic field {\Bin} at $\varphi = 0, 30$, and $90^{\circ}$. {\rhoyxAHE} appears differently below and above the in-plane saturation field {\Bsatin}. (\textit{D}) Magnetic phase diagram of {\ECS} plotted for a {\Bin}-$T$ plane. The color map represents the  derivative of in-plane magnetization $dM_y/dB_y$, plotted together with points determined in the previous reports \cite{ECS_iAHE} and the dashed line separating the AFM and FM phases. (\textit{E}) Corresponding spin configurations in the ground-state A-type AFM, canted AFM, and forced FM states, denoted by symbols ($\bullet$, $\blacktriangle$, $\blacklozenge$) also in (\textit{C}) and (\textit{D}).}
		\label{fig1}
	\end{center}
\end{figure*}

\section*{Results}
\subsection{In-plane AHE depending on magnetic phases}
{\ECS} exhibits substantial but complex in-plane AHE at 2 K \cite{ECS_iAHE}, as shown in Fig. 1\textit{C}. As summarized in Figs. 1\textit{D} and 1\textit{E}, $\mathrm{Eu}^{2+}$ spin magnetic moments ($S = 7/2$) order into an A-type AFM ground state below the N\'eel temperature of {\TN} = 7.5 K, where spins align ferromagnetically within the \textit{ab}-plane and antiferromagnetically along the $c$-axis. This magnetic ordering, belonging to magnetic point group 2/$m$1', breaks $C_3$ rotational symmetry along the out-of-plane $c$-axis, also opening a tiny gap on the degenerate Weyl or Dirac points protected by the $C_3$ symmetry \cite{ECS_Weyl}. The spins gradually cant with increasing in-plane magnetic field, eventually leading to a forced ferromagnetic (FM) phase above the in-plane saturation field of {\Bsatin} = 1.8 T. 
 The in-plane anomalous Hall resistivity {\rhoyxAHE} exhibits pronounced signals with opposite signs at $\varphi = 30^{\circ}$ and $90^{\circ}$, characterized by nonmonotonic behavior in the AFM phase and linear increase in the FM phase. The three-fold (3$\varphi$) angle dependence is clearly observed with nodes at $\varphi = 0^{\circ}, 60^{\circ}, ..., 300^{\circ}$ \cite{ECS_iAHE}. This is in agreement not only with the point group but also with the magnetic point group, since the spins in the canted AFM and forced FM phases also rotate within the plane for the in-plane field rotation \cite{ECS_magnetic}.

\begin{figure*}
\begin{center}
\includegraphics[width=15cm]{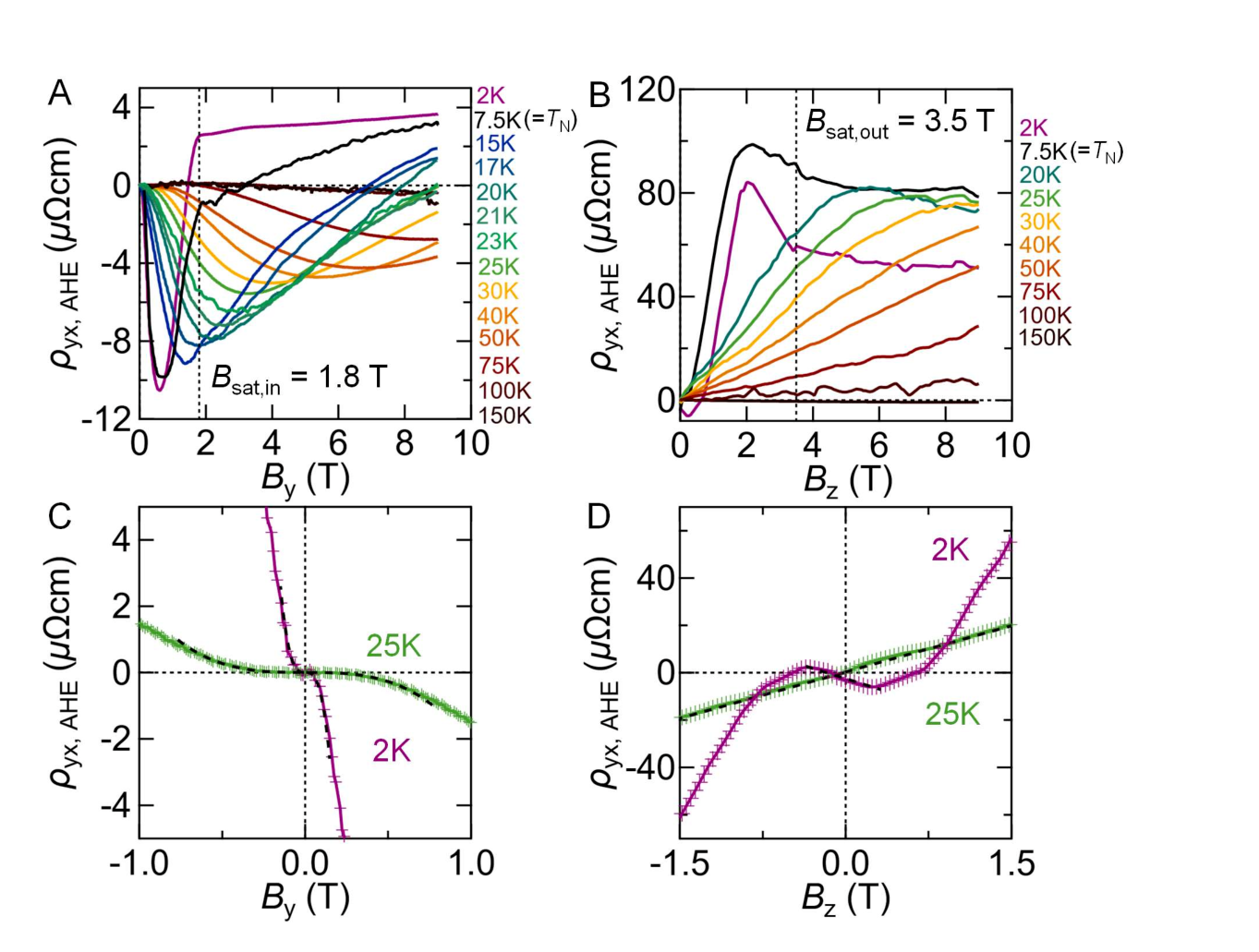}
\caption{Contrasting temperature evolution and field dependence between in-plane and out-of-plane AHE. (\textit{A}) In-plane and (\textit{B}) out-of-plane field sweeps of {\rhoyxAHE}, taken for a {\ECS} film over a wide temperature range well above the N\'eel temperature of {\TN} = 7.5 K. Magnification of the (\textit{C}) in-plane and (\textit{D}) out-of-plane field sweeps measured at $T$ = 2 and 25 K around zero field, which reveal contrasting field dependences, magneto-cubic ({\rhoyxAHE} $\propto$ $B_{y}^3$) and magneto-linear ({\rhoyxAHE} $\propto$ $B_{z}^1$) dependences, respectively.}
\label{fig2}
\end{center}
\end{figure*}

\subsection*{In-plane AHE around zero magnetic field}
Figures 2\textit{A} and 2\textit{B} compare in-plane ($y$ direction) and out-of-plane ($z$ direction) field sweeps of {\rhoyxAHE} taken at various temperature. At 2 K, {\rhoyxAHE} for the in-plane field is roughly, but merely, one order of magnitude smaller than that for the out-of-plane field. With increasing temperature, {\rhoyxAHE} generally decreases in both sweeps and a peak like structure below the saturation field, attributed to nonmonotonic change in the Berry curvature during the magnetization process \cite{ETO_AHE}, gradually shift to higher fields with broadening. As can be seen from the high-temperature data, the initial increase from zero field in the in-plane AHE deviates from a simple linear behavior as in the out-of-plane AHE. 

Figures 2\textit{C} and 2\textit{D} show the magnification of in-plane and out-of-plane  sweeps of {\rhoyxAHE} at 2 and 25 K around zero field. The most important finding is that both {\rhoyxAHE} curves below and above {\TN} for the in-plane field exhibit nonlinear field dependence around zero field, which can be well fitted by a cubic function of the in-plane field. This is in sharp contrast to the linear initial increase from zero field in the case of out-of-plane AHE, though nonmonotonic behavior at finite fields is confirmed in both in-plane and out-of-plane AHE.
Considering the magnetic ordering in the AFM ground state, it is highly nontrivial which of magneto-linear or magneto-cubic dependence dominantly emerges. Nevertheless, clear magneto-cubic dependence is observed not only in the paramagnetic (PM) phase, but also even in the AFM phase. 
This indicates that breaking of the $C_3$ rotational symmetry by the AFM ordering has little effect on in-plane AHE as well as electronic band structures in {\ECS}. Namely, the in-plane AHE is understood as being induced  in the course of the electronic structure evolution from a gapless state with Dirac points to one with split Weyl points.

\begin{figure*}
\begin{center}
\includegraphics[width=15cm]{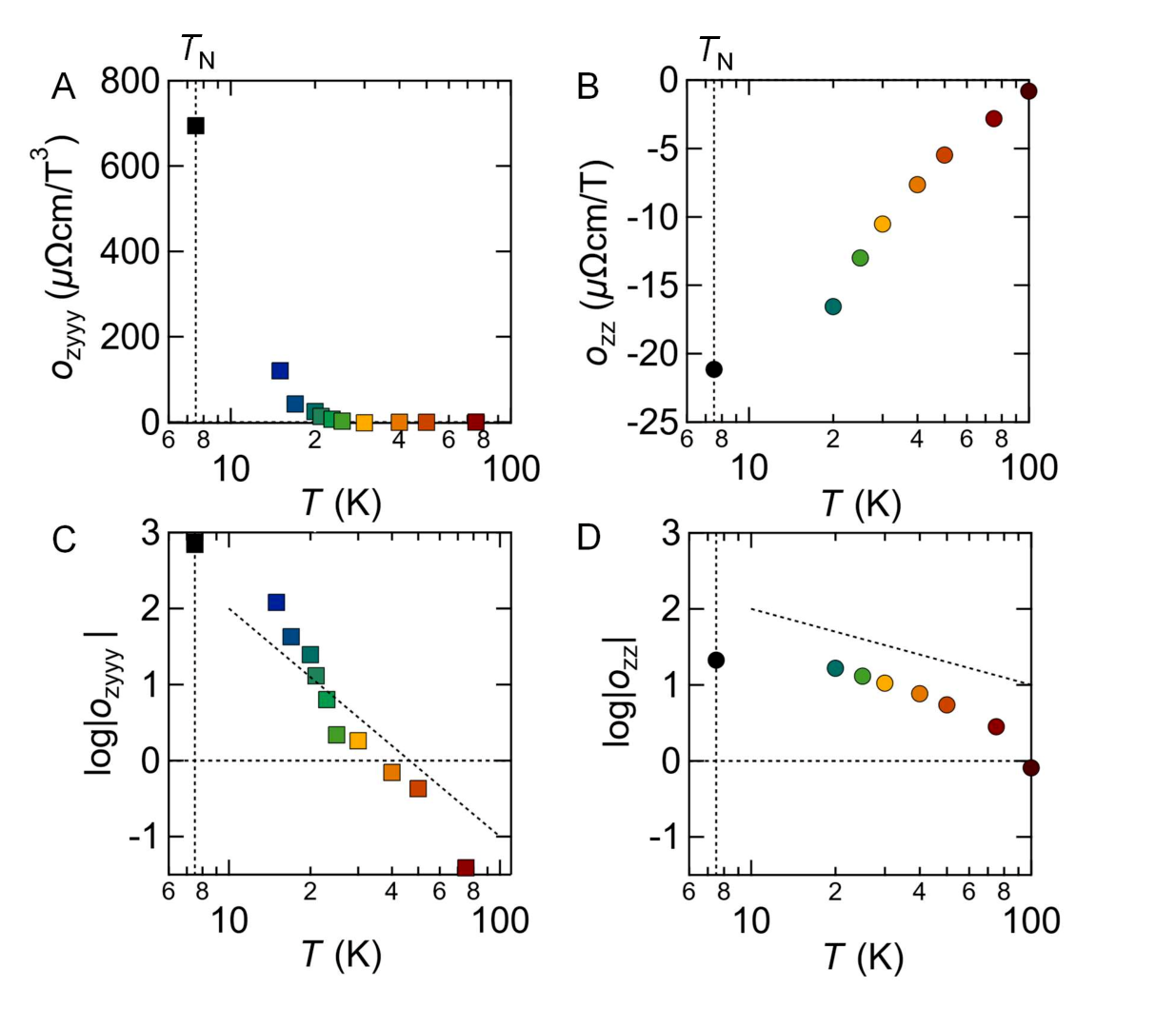}
\caption{Cubic and linear trends in the detailed temperature dependence. Temperature change of the response coefficients (\textit{A}) ${o}_{zyyy}$ and (\textit{B}) ${o}_{zz}$ above {\TN}, extracted by fitting the data shown in Fig. 2 with $\rho_z$(= -{\rhoyx})=${o}_{zyyy}B_y^3$ for in-plane AHE and $\rho_z$(= -{\rhoyx})=${o}_{zz}B_z$ for out-of-plane AHE around zero field, respectively. Temperature change of the logarithm of (\textit{C}) $|{o}_{zyyy}|$ and (\textit{D}) $|{o}_{zz}|$, which clarify contrasting dependences between in-plane and out-of-plane AHE. Namely, dependences on the inverse temperature to the third and first powers (${o}_{zyyy} \propto T^{-3}$ and ${o}_{zz} \propto T^{-1}$) as derived from thermodynamics considerations are roughly confirmed. The dashed line represents decay at the third or first order for a guide to the eye.
}
\label{fig3}
\end{center}
\end{figure*}

Here we introduce a component of the octupolar tensor, ${o}_{zyyy}$, describing the third-order response of $\rho_z$ (= -{\rhoyx}) to the applied in-plane field $B_y$ along the $y$ direction. Namely, the off-diagonal coupling of in-plane AHE in the AFM and PM phases can be formulated as 
\begin{equation}
	\rho_z=o_{zyyy}B_{y}^{3}.
\end{equation}
On the other hand, the diagonal coupling of out-of-plane AHE can be expressed as 
\begin{equation}
	\rho_z=o_{zz}B_{z}.
\end{equation}
In the case where $\rho_z$ is induced via coupling to internal (mainly spin) magnetization $M_y$ along the applied field direction \cite{SRO_iAHE}, instead of Lorentz force by $B_y$, we can also formulate the off-diagonal coupling as
\begin{equation}
	\rho_{z} = \tilde{o}_{zyyy}(aB_{y} + bM_{y})^{3}, 
\end{equation}
with some constants $a$ and $b$. Assuming that $M_y$ reduces with following the Curie-Weiss law above {\TN}, the relation
\begin{equation}
	{o}_{zyyy}=\tilde{o}_{zyyy}\left(a + \frac{bC}{T + |\theta|}\right)^{3}B_{y}^{3}
\end{equation}
can be derived with the Curie constant $C$ and the negative Weiss temperature $\theta$.
Namely, decay of ${o}_{zyyy}$ at higher orders of the inverse temperature, roughly the third order of it, is expected for the in-plane AHE above {\TN}, in contrast to the linear order of it for $o_{zz}$ in the out-of-plane AHE.

Figures 3\textit{A} and 3\textit{B} compare temperature change of the response coefficients $o_{zyyy}$ and $o_{zz}$, extracted by fitting the data around zero field with Eqs. (1) and (2), respectively.
It is evident that $o_{zyyy}$ reduces more rapidly with magnetic field than $o_{zz}$. To clarify this trend further, temperature change of the logarithm of $|o_{zyyy}|$ and $|o_{zz}|$ is plotted in Figs. 3\textit{C} and 3\textit{D}.
They exhibit a general trend that $o_{zyyy}$ and $o_{zz}$ decay roughly following dependences on the inverse temperature to the third and first powers, respectively, although more intricate dependencies including the curvature may appear depending on the values of $a, b, C$, and $\theta$.
This also confirms that the in-plane magnetic field plays an intrinsic role in AHE by altering the band structure and its quantum geometric properties through the induction of in-plane magnetization.

\subsection{In-plane AHE in the forced FM phase}
Having explored the response around zero field, we now move on to the high-field response, especially in the forced FM state. Continuing increase in {\rhoyxAHE} even above the saturation field has been previously observed up to 9 T \cite{ECS_iAHE}. To investigate how this behavior evolves at even higher fields, we performed the high-field measurement up to 24 T. As confirmed in Fig. 4\textit{A}, {\rhoyxAHE} continues to increase linearly with the in-plane field far above the saturation field. This clearly indicates that the dipolar tensor component $o_{zy}$ is finite in the forced FM state and that the magneto-linear contribution, instead of the magneto-cubic contribution, becomes a leading term of the off-diagonal coupling above the saturation field. Namely, breaking of the $C_3$ rotational symmetry by the in-plane FM ordering has significant effect on the band structure by separating the Weyl points along one direction, leading to the dominant magneto-linear dependence in the in-plane AHE. 

Figure 4\textit{B} illustrates schematic change of the band structure under the in-plane field applied along the $y$ direction. In the AFM ground state, a gap opening with the $C_3$ symmetry breaking is very tiny and it can be regarded as hosting Dirac points. So, again, the leading term around zero field is magneto-cubic even in the AFM phase. Upon application of $B_y$, a Dirac point at $\textbf{\textit{k}}_\mathrm{D}$ split into Weyl points, as shown in Fig. 4\textit{B}.
Under the relativistic cross term such as $k_y^{n}\sigma_z$, that splitting occurs not only along the $k_y$ direction but also along the $k_z$ one, leading to finite AHE as proposed in the orbital magnetization model \cite{ECS_iAHE}. With increase of $B_y$, the Weyl points splitting continue to increase. The spin magnetization along the $y$ direction saturates at {\Bsatin}, but the Weyl points splitting continues to increase along the both $k_y$ and $k_z$ directions even above {\Bsatin} in the forced FM state. This leads to continuing change in in-plane AHE and out-of-orbital magnetization, with the magneto-linear leading term  in the FM phase.

\section*{Discussion}
Systematic measurements of in-plane AHE on the {\ECS} (001) plane have demonstrated three key characteristics of the off-diagonal coupling between the applied magnetic field and the Hall vector. First, magneto-cubic dependence of {\rhoyxAHE} dominantly emerges not only in the PM phase but also even in the AFM phase. Second, the off-diagonal component $o_{zyyy}$ shows unconventional decay above {\TN}, roughly depending on the inverse temperature to the third order, in contrast to the first order in $o_{zz}$.
Finally, magneto-linear dependence of {\rhoyxAHE} dominantly emerges 
in the FM phase and it persists up to very high fields. These features are also consistent with the orbital magnetization model \cite{ECS_iAHE}, where the cross term such as $k_y^{n}\sigma_z$ by the spin-orbit coupling (SOC) is essential for the off-diagonal coupling.

This picture is in stark contrast to anomalous Hall antiferromagnets without time-reversal symmetry \cite{altermag_1, altermag_2, altermag_3, altermag_4, altermag_5}, as recently called altermagnets. In this case, time-reversal symmetry is broken even without net magnetization in the absence of SOC, and SOC, required for inducing a finite Hall response, usually induces small but finite spin magnetization perpendicular to the Hall deflection plane \cite{altermag_1}. In in-plane anomalous Hall magnets, on the other hand, SOC is taken into account from the beginning so that a finite Hall response associated with out-of-plane orbital magnetization can emerge even when the spin magnetization lies entirely in the Hall deflection plane.

\begin{figure*}
	\begin{center}
		\includegraphics[width=14cm]{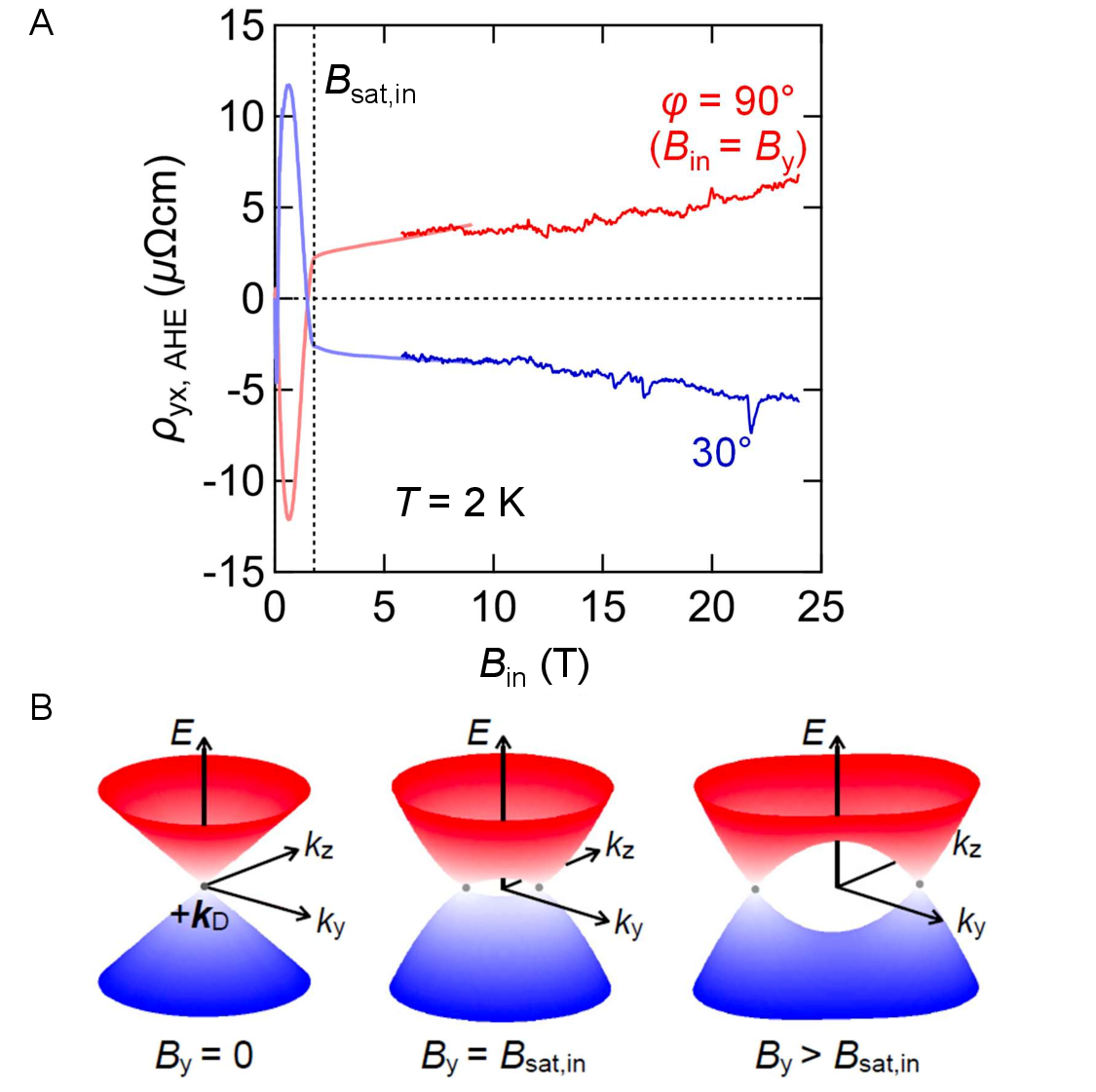}
		\caption{Continuing linear change of in-plane AHE in the forced FM state. (\textit{A}) In-plane field dependence of {\rhoyxAHE}  at $\varphi = 30^{\circ}$and $90^{\circ}$, extended by measuring the same {\ECS} film under high fields up to 24 T as shown by the dark-colored lines. (\textit{B}) Schematic change of the band structure around a Dirac point with increase in the in-plane magnetic field. Continuing increase in {\rhoyxAHE} for $B_y$ above {\Bsatin} can be explained by continuing separation of the Weyl points both along the $k_y$ and $k_z$ directions with the Zeeman splitting of bands.
		}
		\label{fig4}
	\end{center}
\end{figure*}

\vspace{1em}

\section*{Conclusion}
In summary, we have investigated multipolar off-diagonal coupling between $B_y$ and $\rho_{z}$, which realizes in-plane AHE on the {\ECS} (001) principal plane. $\rho_{z}$ around zero field exhibits clear magneto-cubic dependence ($\propto$ $B_{y}^{3}$) not only in the PM state but also in the AFM ground state, regardless of the $C_3$ symmetry breaking by the in-plane AFM ordering. The extracted octupolar off-diagonal component $o_{zyyy}$ is roughly dependent on the inverse temperature to the third ($\propto$ $T^{-3}$) above {\TN}, as expected from thermodynamic considerations. Moreover, $\rho_{z}$ shows magneto-linear dependence ($\propto$ $B_{y}^{1}$) up to very high fields in the forced FM state. These findings, consistent with the orbital magnetization model, may capture key characteristics of the non-trivial off-diagonal coupling in more generic systems. Our demonstration is expected to stimulate further studies and applications focusing on the off-diagonal coupling that has overlooked in the out-of-plane AHE.

\section*{Materials and Methods}
\subsection*{Sample preparation}
(001)-oriented thin film of {\ECS} single-crystalline were epitaxially grown on (111)A CdTe substrates using molecular beam epitaxially in an EpiQuest RC1100 chamber. Molecular beams of Eu, Cd, and Sb were supplied from conventional Knudsen cells. CdTe substrates were etched in 0.01 \% Br$_2$- methanol solution and subsequently at 680$\,^{\circ}\mathrm{C}$ under Cd flux to obtain an atomically smooth surface \cite{ECS_film, CdTe_etching}. The substrate temperature during the growth was maintained at 324$\,^{\circ}\mathrm{C}$. Beam equivalent pressures, monitored using an ionization gauge, were set to $2.0\times 10^{-6}$ Pa for Eu, $5.5\times 10^{-4}$ Pa for Cd, and $7.0\times 10^{-6}$ Pa for Sb.
Only a negligibly small amount of $60^{\circ}$ rotated domains is contained in the {\ECS} thin film, which is
crucial for the observation of in-plane AHE in this system \cite{ECS_film, ECS_iAHE}.
\subsection*{Transport measurement}
The Hall resistivity {\rhoyx} was measured on a Hall bar device using a standard four-probe method. Low-temperature magnetotransport measurements up to 9 T were carried out using a Cryomagnetics cryostat system equipped with a superconducting magnet (CMAG) or Quantum Design Physical Property Measurement System (PPMS), in combination with a one-axis sample rotator.
The magnetic field was rotated within the {\ECS} (001) plane, with azimuthal angle $\varphi$ measured from the [110] direction.
Measurements were conducted by varying the applied magnetic field direction within the ($001$) plane relative to the [$110$] direction. High-field measurements up to 24 T were also performed using a  25 T cryogen-free superconducting magnet
(25 T-CSM) at Institute for Materials Research (IMR).

\vskip\baselineskip
\textit{Note added.} After completing this study, we became aware of a paper by J. Chen \textit{et al}. \cite{B3_cubic}, which reports magneto-cubic in-plane Hall effect on (111) plane of a cubic non-magnetic semimetal.

\vskip\baselineskip
\noindent \textbf{Acknowledgement:} 
We thank F. Kagawa and T. Kurumaji for fruitful discussions.
This work was supported by JSPS KAKENHI Grant Numbers JP23K13666, JP23K03275, JP24H01614, JP24H01654, JP25H00841, JP24K21522, JP23H04868, JP23KK0052, JP22H00109, JP22H04933, and JP23K22447 from MEXT, Japan, by JST FOREST Program Grant Number JPMJFR202N and PRESTO Program Grant Number JPMJPR2452, by Toyota Riken Rising Fellow Program funded by Toyota Physical and Chemical Research Institute, Japan, and by STAR Award funded by the Tokyo Tech Fund, Japan.
This work was partly performed at HFLSM under the GIMRT Program of the Institute for Materials Research, Tohoku University (Proposal No. 202408-HMKPA-0082).
\vskip\baselineskip

\vskip\baselineskip
\vskip\baselineskip
\vskip\baselineskip
\noindent


\clearpage
\end{document}